 \documentclass[a4paper,12pt]{article}

\usepackage{fancyhdr}
\usepackage{xspace}
\usepackage{graphicx}
\usepackage{pslatex}
\usepackage{amsmath}
\usepackage{bbold}
\usepackage{amssymb}
\usepackage[latin1]{inputenc}
\usepackage{txfonts}
\usepackage{color}

 \newcommand{\ecm}{e\,{\rm cm}}

 \newcommand{\hp}{\boldsymbol{\hat{\rm p}}}

 \newcommand{\qp}{\boldsymbol{\rm q}_+}
 \newcommand{\qm}{\boldsymbol{\rm q}_-}
 
 \newcommand{\bq}{\boldsymbol{\rm q}}
 \newcommand{\hbq}{\boldsymbol{\hat{\rm q}}}
 
 \newcommand{\hqp}{\boldsymbol{\hat{\rm q}}_+}
 \newcommand{\hqm}{\boldsymbol{\hat{\rm q}}_-}

\def\nn{\nonumber}

\newcommand{\cO}{\mathcal O}

\def\Re{{\rm Re}}
\def\Im{{\rm Im}}

\def\GeV{{\rm GeV}}

\newcommand{\be} {\begin{equation}}
\newcommand{\ee} {\end{equation}}
\newcommand{\bma} {\begin{math}}
\newcommand{\ema} {\end{math}}
\newcommand{\beqa} {\begin{eqnarray}}
\newcommand{\eeqa} {\end{eqnarray}}

\newcommand{\bc} {\begin{center}}
\newcommand{\ec} {\end{center}}

\newcommand{\simgt}{\hbox{ \raise3pt\hbox to 0pt{$>$}
    \raise-3pt\hbox{$\sim$} }}
\newcommand{\simsm}{\hbox{ \raise3pt\hbox to 0pt{$<$}
    \raise-3pt\hbox{$\sim$} }}
    
\begin{document}

\begin{titlepage}
  \begin{flushright}
    TTK-21-31 \\
  \end{flushright}
  \vspace{0.01cm}
  
  \begin{center}
    {\LARGE \bf Probing the tau electric dipole moment at BEPC-II collider energies } \\
    \vspace{1.5cm}
    {\bf Werner Bernreuther}\,$^{a,}$\footnote{\tt
      breuther@physik.rwth-aachen.de}, 
    {\bf Long Chen}\,$^{a,}$\footnote{\tt longchen@physik.rwth-aachen.de}
    {\bf and  Otto Nachtmann}\,$^{b,}$\footnote{\tt o.nachtmann@thphys.uni-heidelberg.de}
    \par\vspace{1cm}
    $^a$Institut f\"ur Theoretische Teilchenphysik und Kosmologie, \\
    RWTH Aachen University,  52056 Aachen, Germany\\
    $^b$ Institut f{\"u}r Theoretische Physik, Universit{\"a}t Heidelberg, 69120 Heidelberg, Germany
    \par\vspace{1cm}
    {\bf Abstract} \\
    \parbox[t]{\textwidth}
    {\small{ We investigate the prospects of  the search for a nonzero $\tau$ EDM form factor  $d_\tau(s)$ 
     in $\tau$ pair production by $e^+ e^-$   collisions at  BEPC-II collider energies.
   We compute the  expectation values and covariances of simple and optimal $CP$-odd observables for 
  $\tau$-pair production   at $\sqrt{s}=4.18$ GeV and $4.95$ GeV with subsequent decays of $\tau^\pm$ into major leptonic 
   or semihadronic modes. For the $\tau$ decays to two pions and three charged pions
   we take the full kinematic information of the hadronic system into account.
    Applying cuts and  using realistic assumptions on the eventually attainable integrated luminosities at these energies,
    ${\cal L}(4.18) = 3\times 10^4~{\rm pb}^{-1}$ and ${\cal L}(4.95) = 10^4~{\rm pb}^{-1}$, respectively, we find the following. 
     By taking into account
     purely semihadronic and semihadronic-leptonic $\tau^+\tau^-$ decays one can achieve with   
     optimal $CP$-odd observables    the
        1 s.d. sensitivities $\delta \Re d_\tau = 4.5\times 10^{-18} \ecm$ $(5.3\times 10^{-18} \ecm)$
   and $\delta \Im d_\tau = 2.2 \times 10^{-18} \ecm$  $(2.7\times 10^{-18} \ecm)$ at $\sqrt{s}=4.18~\GeV$ $(4.95~\GeV)$.   
    
       }}
    
  \end{center}
  \vspace*{0.7cm}

\end{titlepage}

\setcounter{footnote}{0}
\renewcommand{\thefootnote}{\arabic{footnote}}
\setcounter{page}{1}

%%%%%%%%%%%%%%%%%%%%%%%%%%%%%%%%%%%%%%%%%%%%%%%%%%%%%%%%%%%%%%%%%%%%%%%%%%
\section{Introduction} 
\label{sec:intro}
%%%%%%%%%%%%%%%%%%%%%%%%%%%%%%%%%%%%%%%%%%%%%%%%%%%%%%%%%%%%%%%%%%%%%%%%%%

The experimental investigation of the production and decays of $\tau$ leptons at existing high-luminosity, low-energy $e^+ e^-$ colliders offers, among other issues,
 an  opportunity  to 
 search for new physics beyond the standard model (SM) of particle physics. One aspect of this endeavor is the search for a non-zero 
 $\tau$ electric dipole moment (EDM) that is a signature of $CP$ violation beyond the  Kobayashi-Maskawa  mechanism. 
  As the $\tau$ lepton has a very short lifetime the measurement of its static moments has so far not been possible, but instead, information on its 
  nonstatic EDM form factor\footnote{The acronym EDM  is used in this paper for both the 
  static moment and the form factor at $q^2\neq 0$.}  
  can be extracted, for instance, from the measurement of $CP$-violating spin correlations in $\tau$-pair production in $e^+e^-$ collisions.
 For timelike momentum transfer the $\tau$ EDM form factor $d_\tau(q^2)$ can be a complex quantity. 
    The best existing limits on its real and imaginary parts were obtained 
    by the Belle I Collaboration~\cite{Inami:2002ah} at $q^{2}=(10.58~{\rm GeV})^{2}$:
\begin{align}    \label{Eq.01.03}
-2.2 \times 10^{-17} \, {\ecm} &<\Re~d_{\tau}(q^{2})<4.5\times10^{-17} {\ecm} \, \text{
at }\, 95\% \,{\rm C.L.} \, , \notag \\
-2.5 \times 10^{-17} \, {\ecm} &<\Im~d_{\tau}(q^{2})<0.8\times10^{-17} {\ecm} \, \text{
at }\, 95\% \, {\rm C.L.} \,.
\end{align}
Recently we reconsidered this topic and investigated how the sensitivity to $\Re~d_{\tau}$ and $\Im~d_{\tau}$ can be 
significantly enhanced with suitable simple and optimal
 $CP$ observables \cite{Bernreuther:2021elu}, assuming that high-statistics $\tau^+ \tau^-$ decay data will eventually be collected by the Belle II experiment 
 \cite{Abe:2010gxa,Kou:2018nap} at the KEKB collider. We analyzed also  $d_\tau$ in some SM extensions with $CP$-violating 
  Yukawa couplings that can generate a $\tau$ EDM much larger than the electron EDM. We refer to \cite{Bernreuther:2021elu} for an extensive list of
   references on $\tau$ EDM topics.

Because there is another high-luminosity, low-energy $e^+ e^-$ machine in operation, the Beijing BEPC-II collider, where high-statistics 
 $\tau^+ \tau^-$ production and decay data are being recorded by the BESIII experiment \cite{Ablikim:2019hff}, it seems appropriate and of interest to extend 
  our analysis  \cite{Bernreuther:2021elu} to $e^+ e^-$ c.m. energies relevant for  BESIII. This is what we do in this short paper.
  We consider the reactions 
 \begin{equation}\label{Eq.01.04}
e^{+}+ e^{-} \rightarrow \tau^{+} \; + \; \tau^{-}  \rightarrow {\overline B} \; + \; A 
\end{equation}
 at two $e^+ e^-$ c.m. energies, $\sqrt{s}=4.18~\GeV$ (where the $\tau$-pair production cross section is maximal) and $\sqrt{s}=4.95~\GeV$.
 At these energies the BESIII experiment has already collected a large number of $\tau$-pair events.
 In our analysis we take the major semihadronic and leptonic $\tau$ decays, denoted by $\tau^-\to A$ and $\tau^+ \to  {\overline B}$, 
 of polarized $\tau^\mp$ into account.
 Cuts are applied to the charged leptons and to the charged and neutral mesons in the final state. The SM amplitude for the production of polarized 
 $\tau$-pairs is appended by the contribution of a $CP$-violating $\tau$ electric dipole form factor $d_\tau(s)$ 
 as defined in \cite{Bernreuther:2021elu}
 that is taken to be a complex quantity. The decays of polarized $\tau^\mp$ are modeled at the fully differential level with decay density matrices 
 determined in the SM with intermediate resonances taken into account \cite{Bernreuther:2021elu}.
 We consider suitable simple \cite{Bernreuther:1989kc,Bernreuther:1993nd}
and optimal  \cite{Atwood:1991ka,Davier:1992nw,Diehl:1993br} $CP$ observables for tracing the real and imaginary part of $d_\tau(s)$.
 We compute the expectation values and covariances of these observables  for the c.m. energies and  reactions mentioned above  
 and determine the statistical 1 standard deviation (s.d.) sensitivities, in particular for the cases where i) 
 both $\tau$ leptons decay semihadronically and ii) where  one $\tau$ decays semihadronically and the other one leptonically.

%4-1
%%%%%%%%%%%%%%%%%%%%%%%%%%%%%%%%%%%%%%%%%%%%%%%%%%%%%%%%%%%%%%%%%%%%%
\section{Observables and results}
\label{Sec:04}
%%%%%%%%%%%%%%%%%%%%%%%%%%%%%%%%%%%%%%%%%%%%%%%%%%
 We consider $\tau$ pair production in the reactions \eqref{Eq.01.04} to leading order in the SM supplemented by the 
 $\tau$ EDM form factor. This is adequate for our purpose of computing 
  expectation values of normalized observables. Total cross sections of the reactions \eqref{Eq.01.04} are used only for estimates of corresponding event numbers.
  At the energies of the BEPC-II collider $Z$-boson exchange is negligible.
  
  If one considers events where only one charged 
 particle is measured both in $\tau^-$ and $\tau^+$ decay, respectively, 
  \begin{equation}
 \label{eq:1prong}
 \tau^- \rightarrow a(q_-) + X \, , \qquad \tau^+ \rightarrow {\bar b}(q_+) + X' \, ,
\end{equation}
 the following simple $CP$-odd observables can 
  be applied \cite{Bernreuther:2021elu,Bernreuther:1989kc,Bernreuther:1993nd}:
\begin{eqnarray}
\widehat{T}^{ij} & = &(\hqp-\hqm)^{i}\,\frac{(\hqp\times\hqm)^{j}}{|\hqp\times\hqm|} \, + \, (i\leftrightarrow j) \;,
\label{Eq.04.01} \\
T^{ij} & = & (\qp-\qm)^{i}\,(\qp\times\qm)^{j}  \, + \, (i\leftrightarrow j) \;,
\label{Eq.04.02} \\
\widehat{Q}^{ij}& = & (\hqp+\hqm)^{i}\, (\hqp -\hqm)^{j}+(i\leftrightarrow j) \;,
\label{Eq.04.03} \\
Q^{ij}& = & (\qp + \qm)^{i}\,(\qp -\qm)^{j}-\dfrac{1}{3}\delta^{ij}(\qp^2 - \qm^2)+(i\leftrightarrow j) \;.
\label{Eq.04.04}
\end{eqnarray}
The charged-particle three-momenta ${\bq}_\mp$ in \eqref{Eq.04.01} -- \eqref{Eq.04.04} are defined 
in the $e^+ e^-$ c.m. frame, ${\hbq}_{\pm}={\bq}_{\pm}/|{\bq}_{\pm}|$,
and $i,j\in\lbrace1,2,3\rbrace$ denote the Cartesian vector indices. 
The observables \eqref{Eq.04.01}, \eqref{Eq.04.02} and \eqref{Eq.04.03}, \eqref{Eq.04.04}
are sensitive to $\Re d_{\tau}(s)$ and $\Im  d_{\tau}(s)$, respectively.

For  $e^+ e^-$ c.m. energies considered in this paper (i.e., $\sqrt{s} < 5~\GeV$),
 and using ${\hat d}_\tau(s) = d_\tau(s) \sqrt{s}/e$, $e=\sqrt{4\pi\alpha_{em}}$, the expectation values
 of the observables \eqref{Eq.04.01} -- \eqref{Eq.04.04} are of the form:
  \begin{eqnarray}
   \langle T^{ij}\rangle_{ab} = c_{ab}(s) \, \Re {\hat d}_\tau(s)~ s^{ij} \, , &
   \qquad  \langle \widehat{T}^{ij}\rangle_{ab}=   {\tilde c}_{ab}(s)  \, \Re {\hat d}_\tau(s)~ s^{ij}  \, ,
   \label{eq:exsimT} \\
    \langle Q^{ij}\rangle_{ab} = \kappa_{ab}(s)  \, \,\Im {\hat d}_\tau(s)~ s^{ij} \, , &
   \qquad  \langle \widehat{Q}^{ij}\rangle_{ab}=   {\tilde\kappa}_{ab}(s) \, \Im {\hat d}_\tau(s)~ s^{ij}  \, .
   \label{eq:exsimQ}
  \end{eqnarray}
  Here the labels $a,b$ denote charged particles from $\tau^\mp$ decay, cf. Eq.~\eqref{eq:1prong}.
 
 We evaluate  \eqref{eq:exsimT} and \eqref{eq:exsimQ} for the decay modes  $\tau^-\to \pi^-(q_-)\nu_\tau,~ \rho^-(q_-) \nu_\tau,~ \ell^-(q_-) {\bar\nu}_\ell \nu_\tau,~ (\ell=e, \mu)$ and the corresponding charge-conjugate
  decay modes. The expectation values in  \eqref{eq:exsimT} and \eqref{eq:exsimQ}  are defined by
  \begin{eqnarray}\label{Eq.04.06}
\langle T^{ij} \rangle_{ab} & \equiv & \dfrac{1}{2}\bigl{\lbrace}\langle T^{ij}\rangle_{a\bar{b}}+ \langle T^{ij}
\rangle_{b\overline{a}} \bigr{\rbrace} \nn \\
& = & \dfrac{1}{2}\Bigl{\lbrace}\dfrac{\int d\sigma_{a\bar{b}} T^{ij}}{\int d\sigma_{a\bar{b}}}+
\dfrac{\int d\sigma_{b\overline{a}}
 T^{ij}}{\int d\sigma_{b\overline{a}}}\Bigr{\rbrace}
\end{eqnarray}
 and likewise for the other observables. 
  The  form \eqref{eq:exsimT}, \eqref{eq:exsimQ}
   of the expectation values holds 
    when the phase-space cuts are $CP$-symmetric and rotational invariant. With the $e^+$ beam direction $\hp$ in the $e^+ e^-$  c.m. frame  defining
    the $z$ axis of the coordinate system
  the symmetric traceless tensor $s^{ij}$ is given by
  \begin{equation} \label{eq:defsij}
   (s^{ij})= \frac{1}{2} \left({\hat p}^i {\hat p}^j - \frac{1}{3} \delta^{ij} \right)
   = {\rm diag}\left(-\frac{1}{6}, -\frac{1}{6}, \frac{1}{3} \right) \, .   
  \end{equation}

  We calculate the expectation values of the simple $CP$ observables -- and also of the optimal observables below -- and  the resulting 
  statistical sensitivities to 
  $\Re d_{\tau}(s)$ and $\Im  d_{\tau}(s)$ at the $e^+e^-$ c.m. energies $\sqrt{s}= 4.18~\GeV$ and $4.95~\GeV$.
   The following cuts are applied in the $e^+e^-$ c.m. frame to the transverse momenta and polar angles of 
the pions  and charged leptons from $\tau^\mp$ decay:\footnote{F. Nerling, private communication.}
\begin{equation}\label{eq:cuts}
  p_T > 0.04~{\rm GeV} \, , \qquad |\cos\theta| < 0.93 \, . 
\end{equation}
In the case of the above simple observables we use the reconstructed $\rho$ meson from of $\tau\to \rho\nu_\tau\to 2 \pi \nu_\tau$ as $\tau$-spin analyzer 
and apply the cuts \eqref{eq:cuts} to this particle. However, the cross sections associated with this mode are calculated by applying the cuts \eqref{eq:cuts}
to both the charged and neutral pion from $\rho$ decay. 
At the c.m. energies $4.18$ GeV and $4.95$ GeV we assume the following integrated luminosities:
\begin{equation}\label{eq:intL}
 {\cal L}(4.18) = 3\times 10^4~{\rm pb}^{-1} \, , \quad {\cal L}(4.95) = 10^4~{\rm pb}^{-1} \, .
\end{equation}
Because the diagonal elements of the above traceless symmetric tensor observables in \eqref{eq:exsimT}, \eqref{eq:exsimQ}
are not independent of each other,
 only their $3,3$ components are considered that have the largest expectation values. 

 SM-generated $CP$-violating effects in the flavor-diagonal reaction $e^+e^- \to \tau^+\tau^-$ are extremely 
  tiny. (For a short recent discussion, see,  e.g., \cite{Bernreuther:2021elu}.)
 For the $CP$-symmetric cuts \eqref{eq:cuts}, the 
  SM expectation values of the  above $CP$-odd observables ${T}^{ij}$ and $\widehat{T}^{ij}$, which are also 
   odd under   naive ``time reversal'' $(T_N)$ invariance,
are therefore numerically extremely small  in the diagonal decay channels $a{\bar a}$.
 For non-diagonal decay channels these expectation values can receive also contributions from the absorptive parts of higher-order 
  terms in the SM amplitude, but they can be neglected for our purposes, too.  
 The SM expectation values of ${Q}^{ij}$ and $\widehat{Q}^{ij}$ in the 
 diagonal decay channels $a{\bar a}$ are also negligibly small.
 However, for the non-diagonal decay channels the SM expectation values of the $Q$ tensors become
 as large as  a few $\times 10^{-2}$ in magnitude
  in the presence of cuts \eqref{eq:cuts}. Therefore one should consider in addition to \eqref{Eq.04.03}, \eqref{Eq.04.04}
 the observables  \cite{Bernreuther:2021elu}
 \begin{equation} \label{eq:Qprimevar}
 {Q'}^{ij} = {Q}^{ij} -  \langle  {Q}^{ij} \rangle_0 \, , \qquad  \widehat{Q'}^{ij} = \widehat{Q}^{ij} -  \langle  \widehat{Q}^{ij} \rangle_0 \, .
  \end{equation}
   The label $0$ indicates that the 
   expectation value is computed in the SM.
The SM expectation values of the $Q$ observables in an off-diagonal channel and the respective charge-conjugate channel are 
opposite in sign for $CP$-symmetric cuts:
\begin{equation} \label{eq:opposite}
 \langle  {Q}^{ij} \rangle_{0,a{\bar b}}  =  - \langle  {Q}^{ij} \rangle_{0, b{\bar a}}
\end{equation}
and likewise for $\widehat{Q}^{ij}$. Thus, we have, in view of the definition \eqref{Eq.04.06},
\begin{equation} \label{eq:QQprime}
\langle  {Q'}^{ij} \rangle_{a b} = \langle  {Q}^{ij} \rangle_{a b}
 \end{equation}
 for all decay channels and likewise for  $\widehat{Q}^{ij}$. Thus the values of the coefficients $\kappa_{ab}$ and $\tilde\kappa_{ab}$
 in  \eqref{eq:exsimQ} that we compute for the tensors $Q$ hold also for the tensors $Q'$. 
 That is, the expectation values \eqref{Eq.04.06} of ${Q}^{ij}$ signaling a possible $\tau$ EDM are independent of $\langle Q^{ij} \rangle_0$. 
 But the sensitivity to the $\tau$ EDM attainable with the measurement of this observable must be calculated from the covariance of 
 ${Q'}^{ij}$, cf. \cite{Bernreuther:2021elu}.
 Furthermore we note that the  covariance matrix of the $T$ and $Q'$ tensors is diagonal because the $T$ tensors are $T_N$-odd 
 while the $Q'$ tensors are $T_N$-even.

  As to the sensitivity to the $\tau$ EDM let us use as an example the measurements  of
   $T_{33}$  and $Q'_{33}$ in the decay channels $a\bar{b}$ and $b\bar{a}$.  The resulting
   ideal  1 s.d.  statistical errors of the dimensionful EDM 
  couplings $\Re d_\tau$  and  $\Im d_\tau$ are given by
  \begin{equation} \label{eq:1sdsenTQ}
  \delta \Re d_\tau(s) = \frac{e}{\sqrt{s}}
   \frac{1}{\sqrt{N_{ab}}} \frac{3 \left[\langle T_{33}^2 \rangle_{ab}\right]^{1/2}}{|c_{ab}|} \, ,  \quad
   \delta \Im d_\tau(s) = \frac{e}{\sqrt{s}}
   \frac{1}{\sqrt{N_{ab}}} \frac{3 \left[\langle {Q'}_{33}^2 \rangle_{ab}\right]^{1/2}}{|\kappa_{ab}|} \, .  
  \end{equation}
  The event numbers of the various channels are determined by $N_{aa}= {\cal L}\sigma_{aa}$ and
  $N_{ab}=  {\cal L}\sigma_{ab}$ for $a\neq b$,
  where $\sigma_{ab} =(\sigma_{a {\bar b}} + \sigma_{b{\bar a}})$ denote the respective cross sections.
  In the case of leptonic $\tau$  decays we sum over $\ell=e, \mu$. The cross sections that we use for 
  estimating the event numbers are 
  given below in Table~\ref{tab:optresup}.
 
  Our results for the expectation values and square roots of the variances of  the simple $CP$ observables are listed
  in Tables~\ref{tab:TThatres} and~\ref{tab:QQhatres}
for several one-prong  decays of $\tau^\mp$. In addition, the 1~s.d. sensitivities to the real and imaginary parts of the $\tau$ EDM form factor 
 are given that result from these numbers and from the assumed integrated luminosities \eqref{eq:intL}.
 Our results show that at the energies of the BEPC-II collider the dimensionful  observables $T_{33}$ and $Q'_{33}$ are more sensitive to the real and imaginary 
 part of the $\tau$ EDM than their dimensionless counterparts.
 
 The simple $CP$ observables can be straightforwardly determined by experiment. They require the measurement of the momenta of $ e, \mu$, of charged pions, and 
 also of neutral pions in order to reconstruct the momenta of $\rho^\mp$ in the decay $\rho^\pm \to \pi^\pm \pi^0$, in the  $e^+ e^-$ c.m. frame, i.e., the laboratory frame of the BEPC-II collider.

\vspace{2mm}
\begin{table}[htbp]
\begin{center}
  \caption{Expectation values and square roots of the variances of the observables $T^{ij}$ and $\widehat{T}^{ij}$ 
  defined in \eqref{Eq.04.01}, \eqref{Eq.04.02} and resulting sensitivities to $\Re d_\tau$ assuming the
  luminosities \eqref{eq:intL}.
  For each channel the first row corresponds to the $e^+ e^-$  c.m. energy $\sqrt{s}=4.18\, \GeV$ and the second one to $4.95\, \GeV$. 
    } 
  \vspace{1mm}
  {\renewcommand{\arraystretch}{1.2}
\renewcommand{\tabcolsep}{0.2cm}
\begin{tabular}{ c c c c c c c c}  \hline \hline
 $\tau^-\to$ & $\tau^+\to$ & $c_{ab}$ & $\sqrt{\langle T_{33}^2 \rangle_{ab}}$ & $\delta \Re d_\tau$ & ${\tilde c}_{ab}$ & $\sqrt{\langle \widehat{T}_{33}^2 \rangle_{ab}}$ &  $\delta \Re d_\tau$  \\ 
              &             & $[\GeV^3]$& $  [\GeV^3]$                          & $(\times \, 10^{-17} \ecm)$ &         &    & \ $(\times \, 10^{-17} \ecm)$ \\ \hline\vspace{0.4mm}
$\pi^-\nu$ & $\pi^+ \bar\nu$   &   $ 0.257$   &       $1.038 $  &  $5.50 $     &     $0.182 $    &   $0.894  $    &  $6.70 $ \\ 
           &                   &  $0.616 $          &     $1.835 $ & $6.37 $      &   $0.293 $          &  $0.935  $ & $6.82  $ \\ \\
$\rho^-\nu$ & $\rho^+ \bar\nu$    & $0.041 $       & $0.770 $     &  $11.77$    &   $0.046 $      &  $0.890 $  &   $12.13 $ \\
            &                    & $0.098 $      &  $1.497 $       &  $15.08 $   &    $0.060 $         &  $0.985 $  &  $16.21 $ \\ \\
$\pi^-\nu$ & $\rho^+ \bar\nu$    & $0.103 $       & $0.890 $ &     $5.65 $      &   $0.093 $      & $0.875 $ &    $6.15 $ \\ 
          &                       & $0.247 $       & $1.656 $ &    $6.90 $     &    $0.133  $    &  $0.939  $       & $7.27 $  \\ \\
$\ell^-\nu\bar\nu$ & ${\ell'}^+ {\bar\nu}\nu$   &  $0.021 $  &  $0.436 $ & $8.79 $  &   $0.020 $     & $0.829 $ &   $17.56 $ \\ 
                  &                             & $0.050 $ & $0.769 $   &    $10.23 $ &  $ 0.033$     & $0.878 $        & $17.69 $ \\ \\
$\ell^-\nu\bar\nu$ & $\pi^+ \bar\nu$          & $-0.073$ & $0.627$  & $4.62 $   &  $-0.061 $       &  $0.796 $ &   $7.02 $ \\ 
                  &                         & $-0.176 $ & $1.108 $  & $5.32 $    & $-0.098 $  &  $0.849  $    & $7.32 $ \\ \\
$\ell^-\nu\bar\nu$ & $\rho^+ \bar\nu$      & $-0.029 $ &  $0.568 $   &  $7.14 $  &  $-0.031 $       &  $0.834 $   &  $9.80 $ \\             
                &                       & $-0.071 $ &    $ 1.061$  & $8.57 $   & $-0.044 $      &  $ 0.906$    & $11.81 $  \\ \hline \hline
\end{tabular} }
\label{tab:TThatres}
\end{center}
\end{table}

  \vspace{2mm}
\begin{table}[htbp]
\begin{center}
  \caption{Expectation values of the observables $Q'^{ij}$ and $\hat{Q}^{'ij}$ defined in
   \eqref{eq:Qprimevar} (using \eqref{eq:QQprime} and \eqref{eq:exsimQ}),
  square roots of their variances,   and resulting sensitivities to $\Im d_\tau$ assuming the
  luminosities \eqref{eq:intL}.
 For each channel the first row corresponds to the $e^+ e^-$  c.m. energy $\sqrt{s}=4.18\, \GeV$ and 
 the second one to $4.95\, \GeV$. 
   } 
  \vspace{1mm}
  {\renewcommand{\arraystretch}{1.2}
\renewcommand{\tabcolsep}{0.2cm}
\begin{tabular}{c c  c c c c c c}  \hline \hline
 $\tau^-\to$ & $\tau^+\to$ & $\kappa_{ab}$ & $\sqrt{\langle {Q'}_{33}^2 \rangle_{ab}}$ & $\delta \Im d_\tau$ & ${\tilde\kappa}_{ab}$ & $\sqrt{\langle \widehat{Q'}_{33}^2 \rangle_{ab}}$ &  $\delta \Im d_\tau$  \\ 
              &                & $[\GeV^3]$& $  [\GeV^3]$                          & $(\times \, 10^{-17} \ecm)$ &  &    & \ $(\times \, 10^{-17} \ecm)$ \\ \hline\vspace{0.4mm}
$\pi^-\nu$ & $\pi^+ \bar\nu$  &   $-0.450 $   & $0.914 $  &  $2.77 $     & $-0.224 $ &   $0.703 $    &  $4.28 $  \\ 
           &                  &   $-1.034 $   & $1.346  $  & $2.78  $        & $-0.384  $  & $0.675  $     & $3.76 $  \\ \\
$\rho^-\nu$ & $\rho^+ \bar\nu$ & $-0.197 $ & $0.807 $ &  $ 2.57 $ &   $-0.135 $      &  $0.710  $  &  $3.30 $ \\
            &                  &  $-0.474 $   & $1.295 $ & $2.70 $   & $-0.205 $  &    $0.662  $      &  $3.19 $  \\ \\
$\pi^-\nu$ & $\rho^+ \bar\nu$   & $-0.325 $ &    $0.871 $ &   $1.75 $      & $-0.179 $ & $0.709 $  & $2.59 $  \\ 
           &                   &  $-0.759 $  & $1.343 $        & $1.82 $       & $-0.295 $  & $0.679 $  & $2.37 $   \\ \\
$\ell^-\nu\bar\nu$ & ${\ell'}^+ {\bar\nu}\nu$    & $0.121 $ & $0.579 $  &  $2.02 $   &    $0.074 $     & $0.723  $ & $4.14 $ \\ 
                   &                          &  $0.278  $     & $0.880 $ & $2.11 $    & $0.127$ &  $ 0.709$   & $3.71 $ \\ \\
$\ell^-\nu\bar\nu$ & $\pi^+ \bar\nu$         & $-0.168 $ & $0.817 $  & $2.62 $   &  $ -0.075$   &  $0.733 $ &   $5.26 $ \\ 
                   &                         & $-0.385 $ & $1.260 $   & $2.76 $   & $-0.129 $    & $0.725 $  & $4.75 $ \\ \\
$\ell^-\nu\bar\nu$ & $\rho^+ \bar\nu$        & $-0.038 $ &  $0.730 $  & $7.00 $  &  $-0.030 $ &  $ 0.725$  &  $8.81 $ \\
                  &                          & $-0.089 $ &  $1.190 $  & $7.67 $  & $-0.036 $  & $0.706 $   & $11.25  $   \\ \hline \hline
\end{tabular} }
\label{tab:QQhatres}
\end{center}
\end{table}

 A higher sensitivity to the $\tau$ EDM can be obtained with optimal observables. 
 In addition to the one-prong decays $\tau\to \pi(q) \nu_\tau$ and $\tau \to \ell(q){\nu}_\ell \nu_\tau$ we take now also the decays
 $\tau^-\to \pi^-(q_1) \pi^0(q_2) \nu_\tau$ and $\tau^-\to \pi^-(q_1) \pi^-(q_2) \pi^+(q_3) \nu_\tau$ 
 and the corresponding charge-conjugate decay modes of $\tau^+$ at the fully differential level into account.
 In the following the labels $a,b$ denote a pair of these decays of $\tau^\mp$ into one, two, or three measured particles.
  The optimal observables for tracing $\Re d_{\tau}$ $(\Im  d_{\tau})$ can be constructed in straightforward fashion using the term $\chi^R_{CP}$   $(\chi^I_{CP})$ of the 
   $e^+ e^-\to \tau^+\tau^-$ production density matrix element that is proportional to $\Re d_{\tau}$ $(\Im  d_{\tau})$ and the respective $\tau^- \to a+X$ and $\tau^+ \to {\bar b}+X'$
   decay density matrices $\mathcal{D}^a$ and  $\mathcal{D}^{\bar b}$.  The decay density matrices involve the momenta of the 
    particles $a$ and ${\bar b}$ in the respective $\tau^\mp$ rest frame. One gets
 \begin{equation}
  \cO_{R}^{a\bar{b}} = \frac{{\rm Tr}[\chi^R_{CP} \mathcal{D}^a  \mathcal{D}^{\bar b}]}{{\rm Tr}[\chi_{SM} \mathcal{D}^a  \mathcal{D}^{\bar b}]} \, , \qquad
  \cO_{I}^{a\bar{b}} = \frac{{\rm Tr}[\chi^I_{CP} \mathcal{D}^a  \mathcal{D}^{\bar b}]}{{\rm Tr}[\chi_{SM} \mathcal{D}^a  \mathcal{D}^{\bar b}]} \, ,
 \label{eq:optORI}
 \end{equation}
 where the trace is taken with respect to the spin indices of $\tau^-$ and $\tau^+$.  
 The normalization involves the SM part $\chi_{SM}$ of the $\tau^+\tau^-$ production density matrix. 
 Both observables are $CP$-odd. An extensive discussion of the properties of
  $\cO_{R}^{a\bar{b}}$ and $\cO_{I}^{a\bar{b}}$ and    
  a detailed exposition and the explicit forms of the production and decay density matrices 
    are given in \cite{Bernreuther:2021elu}.
  We define, in analogy to \eqref{Eq.04.06},
 \begin{equation}
 \label{eq:exopto}
  \langle \cO_{R,I}^{a b}  \rangle \equiv \frac{1}{2}[\langle \cO_{R,I}^{a {\bar b}}\rangle + \langle \cO_{R,I}^{b {\bar a}}\rangle] \, .
 \end{equation}

 First, we analyze  for the various decay channels and the phase-space cuts \eqref{eq:cuts} whether or not 
 the optimal $CP$-odd observables \eqref{eq:optORI} may have  a non-zero expectation value in the SM, which can be the case
  for $a \neq b$ (cf. \cite{Bernreuther:2021elu}), 
  and whether the  covariance matrices  have sizable nondiagonal entries. We find that the SM expectation values of $\cO_{R}^{a\bar{b}}$ for all decay channels and those of $\cO_{I}^{a\bar{a}}$ for the diagonal channels are very 
    small and can be neglected for our purposes. For some non-diagonal decay channels the SM expectation values 
    of $\cO_{I}^{a\bar{b}}$ can be as large as 
     a few $\times 10^{-2}$ in magnitude. Therefore we define  \cite{Bernreuther:2021elu}:
     \begin{equation} \label{eq:OOprime}
   {\cO'}_{I}^{a\bar{b}} = \cO_{I}^{a\bar{b}}  -  \langle \cO_{I}^{a\bar{b}} \rangle_0 \, .
   \end{equation} 
   Because $\langle \cO_{I}^{a\bar{b}} \rangle_0 = - \langle \cO_{I}^{b\bar{a}} \rangle_0$ for
    $CP$-symmetric cuts we have, in view of the definition \eqref{eq:exopto},
    \begin{equation}\label{eq:exOOpr}
      \langle {\cO'}_{I}^{a b}\rangle = \langle \cO_{I}^{a b} \rangle \, .
   \end{equation} 
   Moreover, we obtain that $|\langle {\cal O}_R^{ a\bar{b}}  {\cal O'}_I^{a\bar{b}} \rangle_0| < {\rm a~few}\times 10^{-4}$ 
   with numerical uncertainties below $10^{-3}$. That is, the covariance matrix of the above optimal observables 
   is diagonal to good approximation.
   Thus, within the precision of our numerical analysis, we have 
   (cf. Eqs.~(89) -- (93) of \cite{Bernreuther:2021elu}):  
    \begin{equation} \label{eq:expoptO}
   \langle \cO_{R}^{a b}  \rangle = w_{a\bar b}(s)~\Re {\hat d}_\tau(s) \, , \qquad 
   \langle {\cO}_{I}^{a b}  \rangle = \langle {\cO'}_{I}^{a b}  \rangle = \omega_{a\bar b}(s)~\Im {\hat d}_\tau(s) \, .
    \end{equation}

 Using the c.m. energies $\sqrt{s}=4.18\, \GeV$ and $4.95\, \GeV$  and the cuts \eqref{eq:cuts} we obtain the 
 results given in Table~\ref{tab:optresup}
  for the expectation values defined in Eq.~\eqref{eq:expoptO} of the optimal observables and the square roots of their 
  variances
  for the $\tau^\mp$ decays listed above to one, two and/or three measured particles.
   The  1 s.d. statistical sensitivities to $\Re d_{\tau}(s)$ and $\Im  d_{\tau}(s)$ given in this table were 
   computed  in analogy to \eqref{eq:1sdsenTQ} 
    with the listed cross sections and the assumed
    integrated luminosities \eqref{eq:intL}.
 The numbers show that, as expected, the optimal observables  $\cO_R$ and $\cO'_I$ are significantly more sensitive to the $\tau$ EDM
 than the observables $T_{33}$ and ${Q'}_{33}$.  
 Taking into account the full kinematic information on the hadronic system in the $\tau\to 2 \pi \nu_\tau$ 
 and $\tau\to 3 \pi \nu_\tau$ decays, as we do here,
 results in maximal $\tau$-spin analyzing power \cite{Rouge:1990kv,Kuhn:1995nn}, as in the decay $\tau\to  \pi \nu_\tau$;
 i.e., the   values of the coefficients   $w_{a\bar b}(s)$ and $\omega_{a\bar b}(s)$ are equal for the respective channels.
  However, this holds only if no cuts are applied.
 The application of cuts to the various $n$-meson final states leads to slight distortions as the 
 numbers in Table~\ref{tab:optresup} show.

   \vspace{2mm}
\begin{table}[htbp]
\begin{center}
  \caption{ Sensitivities to $\Re d_\tau$ and $\Im d_\tau$ from the expectation values \eqref{eq:expoptO} of the optimal 
   observables $\cO_{R}^{a{\bar b}}$ and ${\cO'}_{I}^{a{\bar b}}$ assuming the
  luminosities \eqref{eq:intL}.
  For each channel the first row corresponds to the $e^+ e^-$  c.m. energy $\sqrt{s}=4.18\, \GeV$ 
  and the second one to $4.95\, \GeV$. 
    In the case of $a\neq b$ the  cross section value $\sigma_{a b} = (\sigma_{a {\bar b}} + \sigma_{b {\bar a}})$. For the modes 
    where leptons 
   are involved the cross section is the sum of the respective cross sections for $\ell =e$ and $\mu$; in particular $\sigma_{\ell\ell'}$
    is the sum of the diagonal and nondiagonal channels.
   }  
  \vspace{1mm}
  {\renewcommand{\arraystretch}{1.2}
\renewcommand{\tabcolsep}{0.2cm}  
\begin{tabular}{c c c c c c c c c}  \hline \hline
 $\tau^-\to$ & $\tau^+\to$ &  $\sigma_{ab}$ & $w_{a{\bar b}}$   & $\sqrt{\langle (\cO_{R}^{a{\bar b}})^2 \rangle_0}$ & $\delta \Re d_\tau$ & ${\omega}_{a{\bar b}}$ & $\sqrt{\langle ({\cO'}_{I}^{a{\bar b}})^2 \rangle_0}$ &  $\delta \Im d_\tau$  \\ 
             &             &   $[{\rm pb}]$ &                   &       & $(\times \, 10^{-17} \ecm)$ &         &                   & \ $(\times\, 10^{-17} \ecm)$ \\ \hline\vspace{0.4mm}
$\pi^-\nu$ & $\pi^+ \bar\nu$             &  $35.97  $     &  $0.047 $ &    $0.218 $  &  $2.11 $    &     $0.138 $    &   $0.372 $    &     $1.22 $  \\ 
        &                                  &  $31.26  $   &  $0.085 $   &  $0.292 $     &   $2.45 $  &    $0.235 $      &  $0.485 $   &    $1.47 $      \\ \\
$\pi^-  \pi^0\nu$ & $\pi^+ \pi^0 \bar\nu$  &  $170.13 $ & $0.049 $       &  $0.221  $     &  $0.94 $ &   $0.142 $      &  $0.377 $  & $0.55 $      \\
                  &                        &  $146.72  $ & $0.090$       &   $0.300 $     &   $1.10 $ &  $0.248 $      &   $0.498 $ &  $0.66 $  \\ \\
$\pi^-\pi^-\pi^+\nu$ & $\pi^+ \pi^+\pi^-\bar\nu$  &  $19.51$ & $0.050 $       & $0.224 $ &  $2.76$ &   $0.145 $ &  $0.381 $  &     $1.62 $ \\
                     &                            &  $16.88  $ &  $0.093 $  & $0.304$  &  $3.17  $ &  $0.258 $  &   $0.508 $  &  $1.91  $     \\ \\
$\pi^-\nu$ & $\pi^+ \pi^0 \bar\nu$              &  $156.22 $ & $0.048 $ & $0.220 $ &   $ 1.00$ &   $0.141 $   & $0.375 $ &   $ 0.58$  \\ 
           &                                    &  $135.00  $ & $0.088 $ & $0.297 $  &  $1.16 $  &   $0.243 $  &  $0.492 $  &   $0.69 $               \\ \\
$\pi^-\nu$ & $ \pi^+ \pi^+\pi^-\bar\nu$        &  $52.83 $  & $0.049 $     & $0.222 $     & $1.70 $      &   $0.143 $ &  $0.378 $  &  $0.99 $   \\
             &                                 &  $45.59  $  &  $0.090 $     & $0.300 $ & $1.96 $  &    $0.250 $  &   $0.499 $   &    $1.18 $   \\ \\
$\pi^-\pi^0\nu$ & $ \pi^+ \pi^+\pi^- \bar\nu$  &  $115.18 $ & $0.050 $    & $0.223  $     & $1.13 $      &   $0.144  $    &  $0.380 $  & $0.67 $  \\
                &                             &  $99.41 $  &  $0.092 $ &   $0.303 $  &   $1.32  $      & $0.254 $  &  $0.504 $ & $0.79  $   \\ \\
$\ell^-\nu\bar\nu$ & ${\ell'}^+ {\bar\nu}\nu$  &  $372.47  $ & $0.002 $  & $0.045 $ & $3.18 $  &  $0.028 $  & $0.166 $ &   $0.84 $ \\ 
                  &                            &  $323.30  $ &  $0.004 $ & $0.059 $ &  $3.27 $ &  $0.045 $  &  $0.211 $ &  $1.04 $  \\ \\
$\ell^-\nu\bar\nu$ & $\pi^+ \bar\nu$        &  $231.15  $  & $0.010 $  & $0.097 $ & $1.74 $ &    $0.078 $  &  $0.278 $  &  $0.64 $ \\ 
                   &                        &  $200.54  $  & $0.017 $  & $0.129 $ & $2.14 $  &   $0.127 $   &  $0.356 $ &  $0.79 $  \\ \\
$\ell^-\nu\bar\nu$ & $\pi^+\pi^0 \bar\nu$   &  $503.25  $ & $0.010 $ &  $0.100 $   & $1.21 $ & $0.078 $  &  $0.279 $  &  $0.43 $    \\
                   &                        &  $434.88  $  &  $0.017 $ &  $0.131 $ &  $1.47 $ &  $0.129 $  &  $0.359 $  &  $0.53 $     \\ \\
$\ell^-\nu\bar\nu$ & $ \pi^+ \pi^+\pi^-\bar\nu$  &  $170.28  $ & $0.010 $  & $0.100 $   & $2.08 $ & $0.079 $  &   $0.281 $ &  $ 0.74 $  \\
                  &                              &  $147.00  $ & $0.018 $  & $0.133 $   & $2.43 $  &  $0.132 $  & $0.364 $   & $0.91 $        \\ \hline \hline
\end{tabular} }
\label{tab:optresup}
\end{center}
\end{table}

Assuming uncorrelated errors, we can add in quadrature
the statistical errors  of $\Re d_\tau$ and $\Im d_\tau$ 
 listed in Tables~\ref{tab:TThatres},~\ref{tab:QQhatres}, 
and~\ref{tab:optresup} for the various channels,
\begin{equation} \label{dredab}
 \delta\Re d_\tau  = \left(\sum\limits_{ab} \frac{1}{\left(\delta\Re d_\tau\right)^2_{ab}} \right)^{-1/2} \, ,
\end{equation}
and analogously for $\delta\Im d_\tau$. The results of these quadratures are given in Table~\ref{tab:EDMerrup}. 
 The sensitivity to 
 $\Re d_\tau$  is improved by a factor of about 6 with the optimal observable $\cO_R$ as compared to using  
 $T_{33}$ while  the improvement of the 
  sensitivity to $\Im d_\tau$ by using $\cO'_I$  is slightly smaller.

   \vspace{2mm}
\begin{table}[htbp]
\begin{center}
  \caption{Ideal 1 s.d. statistical errors on $\Re d_\tau$ and $\Im d_\tau$ that result from adding the 
   respective uncertainties given in  Tables~\ref{tab:TThatres},~\ref{tab:QQhatres}, and~\ref{tab:optresup}  in quadrature.
  The first row corresponds to the $e^+ e^-$  c.m. energy $\sqrt{s}=4.18\, \GeV$ and the 
   second one to $4.95\, \GeV$.  
} 
 \vspace{1mm}
  {\renewcommand{\arraystretch}{1.2}
\renewcommand{\tabcolsep}{0.2cm}
\begin{tabular}{c c c c  c c} \hline \hline
$\delta\Re d_\tau \; [\ecm]$ &  &  & $\delta\Im d_\tau \; [\ecm]$&  &  \\ \hline 
 $\langle T_{33}\rangle_{ab}$ & $\langle\widehat{T}_{33}\rangle_{ab}$ &$\langle \cO_{R}^{a b}  \rangle$ &
 $ \langle Q'_{33}\rangle_{ab}$    &$ \langle \widehat{Q'}_{33}\rangle_{ab}$ & $\langle {\cO'}_{I}^{a b}  \rangle$ \\
 $  2.6 \times 10^{-17}$ & $ 3.3 \times 10^{-17}  $ &  $ 4.5  \times 10^{-18} $ &
  $  1.0\times 10^{-17} $ &   $  1.6\times 10^{-17}$ & $ 2.1\times 10^{-18} $ \\
 $  3.0 \times 10^{-17}$  & $  3.7 \times 10^{-17}$  & $ 5.3 \times 10^{-18}$ & $ 1.0 \times 10^{-17}$ & $  1.5\times 10^{-17}$& 
  $  2.6 \times 10^{-18}$  \\ \hline \hline
 \end{tabular} }
\label{tab:EDMerrup}
\end{center}
\end{table}

 \vspace{2mm}
\begin{table}[htbp]
\begin{center}
  \caption{Ideal 1 s.d. statistical errors on $\Re d_\tau$ and $\Im d_\tau$ that result from adding in quadrature the 
   respective uncertainties given in Table~\ref{tab:optresup} attainable with the optimal observables $\cO_{R}^{a b}$ 
     and ${\cO'}_{I}^{a b}$    in the semihadronic decays $( h h)$
   and in the semihadronic and semihadronic-leptonic $( h h + h \ell)$ decays of $\tau^+ \tau^-$.
  The respective first row corresponds to the $e^+ e^-$  c.m. energy $\sqrt{s}=4.18\, \GeV$ and the 
   second one to $4.95\, \GeV$. 
   } 
 \vspace{1mm}
  {\renewcommand{\arraystretch}{1.2}
\renewcommand{\tabcolsep}{0.2cm}
\begin{tabular}{c c c } \hline \hline
                               & $\delta\Re d_\tau \; [\ecm]$ &  $\delta\Im d_\tau \; [\ecm]$  \\ \hline 
 $ h h:$    &  $  5.3 \times 10^{-18}$         &       $ 3.1 \times 10^{-18} $ \\  
            &   $ 6.1  \times 10^{-18}$        &    $ 3.7  \times 10^{-18} $  \\ \\
 $ h h + h \ell:$  &  $  4.5 \times 10^{-18}$   & $  2.1 \times 10^{-18} $  \\
                   &  $  5.3 \times 10^{-18}$     &   $  2.7 \times 10^{-18} $  \\ \hline \hline
 \end{tabular} }
\label{tab:EDMerl}
\end{center}
\end{table}

     The optimal observables involve the momenta of charged leptons and/or charged and neutral pions from $\tau^\pm$ decay in the 
     respective $\tau^\pm$ rest frame. If both $\tau^+$ and $\tau^-$ decay semihadronically or 
      if one of the $\tau$ leptons decays semihadronically and 
      the other one to either $e$ or $\mu$, the rest-frame momenta of these final-state particles can be reconstructed from 
      their measured momenta in the laboratory frame and kinematic constraints, using the method of \cite{Kuhn:1993ra};
       see the corresponding discussion in Sec.~5 of  \cite{Bernreuther:2021elu}.
     If both $\tau$ leptons decay
        leptonically the determination of their momenta is not possible in an unambiguous way.
        
      Discarding the results for the $\ell \ell'$ channels  
        and adding  in quadrature  the statistical errors  of $\Re d_\tau$ and $\Im d_\tau$
        listed in  Table~\ref{tab:optresup}  where i) both $\tau$'s decay semihadronically 
        and ii) where the semihadronic-leptonic decays of $\tau^+ \tau^-$ are added to the purely semihadronic events 
        we obtain the 
         1 s.d. errors given in Table~\ref{tab:EDMerl}.
        Comparing the numbers in this table with those in  Table~\ref{tab:EDMerrup} shows that restriction to these
      $\tau^+ \tau^-$ decays does not significantly decrease the sensitivity to $\Re d_\tau$ and $\Im d_\tau$.

%%%%%%%%%%%%%%%%%%%%%%%%%%%%%%%%%% 
\section{Conclusions}
\label{sec:concl}
%%%%%%%%%%%%%%%%%%%%%%%%%%%%%%% 

We found that the prospects are promising of improving the sensitivity in  the search for a nonzero $\tau$ EDM at the BEPC-II collider
 compared  to existing upper bounds on $\Re d_\tau$ and $\Im d_\tau$.
 We considered $\tau^+ \tau^-$ production and decay at the $e^+e^-$ c.m. energies $\sqrt{s}=4.18~\GeV$ and $4.95~\GeV$.
At these energies the BESIII experiment has already recorded a large number of $\tau$-pair events.
With realistic phase-space cuts on the measured final-state particles and realistic assumptions on the eventually attainable 
 event numbers we found that by taking into account
     purely semihadronic and semihadronic-leptonic $\tau^+\tau^-$ decays one can obtain with the  optimal $CP$-odd observables 
     defined above  the
       1 s.d. sensitivities $\delta \Re d_\tau = 4.5\times 10^{-18} \ecm$ $(5.3\times 10^{-18} \ecm)$
            and $\delta \Im d_\tau = 2.1 \times 10^{-18} \ecm$  $(2.7\times 10^{-18} \ecm)$ at $\sqrt{s}=4.18~\GeV$ $(4.95~\GeV)$.
            Assuming that the EDM form factor $d_\tau(s)$ varies only moderately with the c.m. energy in the interval
            $4.1~\GeV < \sqrt{s} < 5~\GeV$,  we can, for instance for the $hh + h\ell$ case, combine the sensitivities 
            calculated for the two values of $\sqrt{s}$ listed in Table~\ref{tab:EDMerl}.
            We then get the 2 s.d. sensitivities 
           \begin{equation} \label{eq:sens2sd}
            \delta\Re d_\tau = 6.9 \times 10^{-18}~\ecm \, , \qquad \delta\Im d_\tau = 3.3 \times 10^{-18}~\ecm
           \end{equation}
       obtainable with the BESIII experiment with the integrated luminosities \eqref{eq:intL}.
         Since 2 s.d. corresponds roughly to a $95\%$ confidence-level interval we can compare     
           these numbers with the best limits \eqref{Eq.01.03} available to date.
            This shows that improvement of the precision of $\Re d_\tau$ and $\Im d_\tau$ 
             by a factor of around 5  should be feasible by the BESIII experiment.

\section*{Acknowledgments}
The authors thank Frank Nerling  for discussions
 and correspondence. The work of L.C. was supported by the Deutsche Forschungsgemeinschaft under Grant  No.
 396021762-TRR 257.

 \pagebreak
 \newpage

%%%%%%%%%%%%%%%%%%%%%%%%%%%%%%%%%%%%%%%%%%%%%%%%%

\end{document}